\documentclass[preprint,aps,showpacs]{revtex4-1}

\usepackage{graphicx}
\usepackage{dcolumn}
\usepackage{amsmath}
\usepackage{epstopdf}
\usepackage{booktabs}

\begin{document}

\preprint{}

\title[Short Title]{Comprehensive modeling of Joule heated cantilever probes}

\author{M. Spieser$^2$, C. Rawlings$^{1,3}$, E. L\"ortscher$^1$, U. Duerig$^1$, and A.W. Knoll$^1$}
\email{ark@zurich.ibm.com}
\thanks{$^3$ Now at SwissLitho AG, 8805 Z\"urich, Switzerland}

\affiliation{%
$^1$ IBM Research-Zurich, 8803 R\"uschlikon, Switzerland}%

\affiliation{%
$^2$ SwissLitho AG, 8805 Z\"urich, Switzerland}%


\date{\today}

\begin{abstract}
The thermo-electrical properties of a complex silicon cantilever structure used in thermal scanning probe lithography are modeled based on well established empirical laws for the thermal conductivity in silicon, the electrical conductivity in the degenerate silicon support structure, and a comprehensive physical model of the electrical conductivity in the low-doped heater structure. The model calculations are performed using a set of physically well defined material parameters and finite element methods to solve the coupled thermal and electrical diffusion equations in the cantilever. The material parameters are determined from a non-linear regression fit of the numerical results to corresponding measured data which also includes Raman measurements of the heater temperature. Excellent agreement between predicted and measured data in the absence of air cooling is obtained if a tapered doping profile in the heater is used. The heat loss through the surrounding air is also studied in a parameter free three-dimensional simulation. The simulation reveals that the heater temperature can be accurately predicted from the electrical power supplied to the cantilever via a global scaling of the power in the power-temperature correlation function which can be determined from the vacuum simulation.
\end{abstract}


\maketitle

\section{Introduction}

The use of heated atomic force microscopy (AFM) tips for the study of thermal properties \cite{majumdar1993thermal} and for surface patterning \cite{mamin1992thermomechanical} with nanometer scale lateral resolution was pioneered more than 20 years ago and has seen a tremendous research activity ever since. A significant technical step forward was achieved by the development of all Si micromachined heatable AFM probes \cite{chui1997micromachined}. This development enabled the reproducible batch fabrication of such probes similar to standard AFM cantilevers. In essence, the heatable probe consists of a U-shaped two terminal structure made of highly doped Si legs providing the electrical contact to the low doped end section which acts as a resistive heater element for the Si tip. Owing to the micro-meter dimension of the probes fast thermal response times on the order of micro-seconds are achieved. The fast thermal response of such micro cantilevers spurred a significant research effort exploring data storage applications based on the thermal embossing of a polymeric storage medium \cite{chui1996low}. The concept was later expanded by exploiting the scaling provided by the parallel operation of a large number of probes in a cantilever array to achieve competitive read/write data rates \cite{vettiger2002millipede}. More recently, heated cantilever probes have become commercially available and are now increasingly used as sensitive nanoscale thermometers \cite{gomes2015scanning} or as point heat sources to perform chemical modifications on surfaces \cite{garcia2014advanced}. The chemical modification is used for a direct conversion of precursor materials for the chemical functionalization \cite{szoszkiewicz2007high,carroll2013fabricating}, the control of surface charge \cite{doi:10.1021/acs.langmuir.6b03471}, the fabrication of functional electronic \cite{wei2010nanoscale,ADMA:ADMA201202877}, magnetic \cite{albisetti2016nanopatterning}, or ferroelectric \cite{kim2011direct} structures. Thermal scanning probe lithography (tSPL) has emerged as a new technologically promising application. Here, the thermal stimulus induces either a cross-linking reaction in a standard resist material \cite{basu2004scanning} or it causes polymeric resist materials to spontaneously decompose into volatile moieties thereby creating a positive tone lithographic pattern \cite{hua2006nanolithography,gotsmann2006exploiting,pires2010nanoscale,knoll2010probe}. The prediction of the tip temperature from the applied electrical power is one of the critical issues in all of these applications. The problem has been addressed mostly by means of numerical simulations of the thermo-electrical characteristics of U-shaped thermal probes \cite{lee2007thermal,nelson2007temperature}. However, the thermal probes used in our own tSPL work are substantially more complex, see Figure \ref{figure-1}, and published results cannot be readily carried over. Moreover, we feel that a scholarly presentation of the subject, in particular with regard to the description of the thermal and electrical materials parameters, will be beneficial to other researchers working in the field. The other subtle and non-trivial, yet important, aspect discussed in this paper is the thermal coupling to the surrounding air, a subject which has been scarcely discussed from a fundamental physics based modeling point of view.

The paper is organized as follows: The methods and materials parameters used for the thermo-electric modeling are thoroughly discussed in Section II. Special focus is placed on the description of the low doped heater section of the tip. In particular, we introduce a tapered doping profile to account for dopant diffusion occurring during high temperature annealing steps in the manufacturing process of the cantilevers. The tapered profile is crucial to obtain physically meaningful fit results and good agreement of the modeled thermo-electrical cantilever characteristics with experimental data. In Section III we discuss the Raman measurements of the heater temperature which we can determine with an error margin of $\pm \, 5$ K at room temperature rising to $\pm \, 25$ K above 650 K. We use the Raman data as a benchmark test for the predictive power of our simulations. The simulation results for a cantilever operating in vacuum, viz. at negligible heat coupling to air, are presented in Section IV. Despite the sophisticated model used in the simulations the error of the predicted heater temperature was significant, up to 100 K, above 850 K using merely the current-voltage characteristics as input parameter for the fit. However, we show that excellent prediction of the heater temperature within the calibration error of the Raman measurement is obtained using the electrical power supplied to the cantilever as parameter in a empirical temperature-power correlation determined from the model fit including electrical and Raman data. In Section V we elaborate on the thermal coupling of the cantilever to surrounding air. Based on a simple physical argument we show that the coupling efficiency strongly depends on the geometry of the heater and it can reach values as high as $10^4$ W m$^{-2}$K$^{-1}$ in micrometer size structures. Using the fit parameters from the vacuum simulation for the material parameters and published data for the thermal conductivity of air we obtain also for a cantilever in air good agreement between measured and simulated current-voltage characteristics as well as excellent predictability of the heater temperature from the temperature-power correlation established in the vacuum case.

\section{Electro-thermal modeling}

\begin{figure}[!] \centering
\includegraphics[width=50mm]{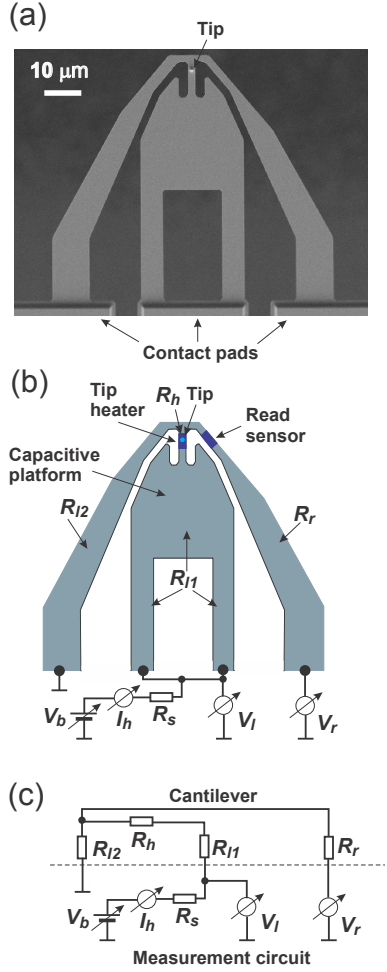}\\
\caption{\footnotesize (a) SEM image of the thermal probes investigated in this study. (b) Schematic of the probe outline and electrical measurement set-up. Low doped heater elements are shown in dark blue whereas the remaining structure (gray) is highly doped to provide a low electrical resistance contact to the bonding pads. Symbols $R_h$, $R_{l1}$, $R_{l2}$, and $R_r$ refer to the resistive elements of the simplified equivalent circuit shown in (c). (c) Simplified equivalent circuit of the cantilever structure and measurement circuit. The high doped leg sections connecting the heater element, $R_h$, to the electrical measurement set-up are represented by the resistors $R_{l1}$ and $R_{l2}$, respectively. The voltage drop occurring over $R_{l2}$ is sensed via the read sensor terminal represented by $R_r$. The series resistor $R_s$ in the external circuit is required in order to obtain a single valued function for the heater current, $I_h$, versus applied bias, $V_h$, when the heater temperature exceeds the inversion temperature of the low doped section giving rise to a negative differential resistance of the overall device.} \label{figure-1}
\end{figure}

Fig. \ref{figure-1} shows the three-legged thermal probe design, also termed cantilever, used in this study. Such cantilevers have been used in our own tSPL work over the past few years. The cantilevers are fabricated from a phosphorous n-doped [100] oriented silicon-on-insulator (SOI) wafer with a nominal doping concentration of 3.3$\times 10^{17}$ cm$^{-3}$ and the cantilever axis points along the [110] direction. The thickness of the cantilever is nominally 400 nm. The fabrication details can be found in Refs\cite{despont2000vlsi,drechsler2003cantilevers}. The structure comprises a resistive heater for the tip which is positioned such that it aligns with the hottest spot of the heater during operation and second heater element positioned on the right arm for thermo-electric height sensing \cite{vettiger2002millipede,durig2005fundamentals}. The heater elements, shown in dark blue in Fig. \ref{figure-1}(b), constitute in essence Si resistors with a doping concentration of the SOI wafer. The heater elements are contacted by the Si support structure which is highly doped at nominally 2.2$\times 10^{20}$ cm$^{-3}$ by means of phosphor ion implantation followed by a 30' thermal activation at 1150$^\circ$ C.

In this study we are only concerned with predicting the steady state tip temperature as a function of the bias potential, $V_b$ which energizes the tip heater using a series resistor, $R_s$, and the heater current $I_h$ is returned to ground via the left lever arm, see Fig. \ref{figure-1}(b). The task is to solve two coupled diffusion equations for the temperature and voltage distributions in the cantilever. The temperature distribution in the steady state is given by
\begin{equation}
\nabla_\bullet \left ( \kappa(T) \, \nabla T(x,y) \right ) = \dot{q}
\label{equ1}
\end{equation}
where $\kappa$ is the thermal conductivity in Si which only depends on the temperature $T$, see below, and $\dot{q}$ is the density of the dissipated electrical power. Similarly we have for the voltage distribution
\begin{equation}
\nabla_\bullet \left ( \sigma(x,y,T) \, \nabla V(x,y) \right ) = 0
\label{equ2}
\end{equation}
and the coupling term due to Joule heating is given by
\begin{equation}
\dot{q} = \sigma(x,y,T) \, (\nabla V {\cdot} \nabla V)
\label{equ3}
\end{equation}
where the electrical conductivity $\sigma$ depends on the temperature as well as on the local doping concentration and thus also becomes a function of the planar coordinates. The boundary conditions for the temperature equation are $T$ = 293 K at the end sections of the lever. For the voltage equation we have $V$ = 0 at the grounded terminal, $V \, = \, V_b - I_h R_s$ at the center terminals, and $\nabla V {\cdot} \vec{n}_x \, = \, 0$ at the $V_r$ terminal where $\vec{n}_x$ denotes the unit vector along the lever axis. The total heater current $I_h$ is obtained by integrating the current density $j_x \, = \, \sigma \, (\nabla V {\cdot} \vec{n}_x)$ either at the center terminals or at the grounded left terminal over the corresponding lever cross sections.

The predictive power of the simulation critically depends on how well the materials parameters $\kappa$ and $\sigma$ describe the physical behavior of the cantilever. The problem is accentuated by the fact that the thermal conductivity depends on the details of the phonon scattering mechanism due to impurities and the nearby boundary in thin Si structures \cite{asheghi2002thermal}. Likewise, the electrical conductivity strongly depends on the doping concentration and electron scattering in the thin Si structure. These effects cannot be predicted a priori with sufficient accuracy. Therefore we are forced to introduce a set of fit parameters henceforth denoted by bold symbols in the materials equations.

The thermal conductivity in high quality single crystal Si, $\kappa_{Si}$, is taken from Table IV in Ref.\cite{fulkerson1968thermal} and it has been shown that $\kappa_{Si}$ does not significantly depend on the doping concentration in the temperature, 300 K - 1200 K, and doping regime, 10$^{17}$ - 10$^{20}$ cm$^{-3}$, considered here \cite{stranz2013thermoelectric}. In essence $1/\kappa_{Si}$ increases linearly with temperature with a distinct increase of the slope at 680 K. With reference to $\kappa_{Si}$ we write
\begin{eqnarray}
\kappa (T) & = & {\bf c_\kappa} \, \kappa_{Si}(T)  \label{equ4} \\
\frac{1}{\kappa_{Si} (T)} & \simeq & 1 \, {\rm W^{-1} cm K} \, \left ( 0.707 + 3.15 \times 10^{-3} \, (T-300 \, {\rm K}) +  \right . \nonumber \\
& & \left . + 5.25 \times 10^{-4} \, (1+\tanh(T-680 {\rm K})) \,  \, (T-680 {\rm K}) \right ) \nonumber \,\,\,
\end{eqnarray}
where the $\tanh$ term provides an approximation for the step wise increase in the slope at 680 K. The scale parameter ${\bf c_\kappa}$ accounts for the reduction of the thermal conductivity due to increased phonon scattering in our structure. ${\bf c_\kappa}$ is expected to be less than 1 which provides a first sanity check for the model.

The electrical conductivity in the low doped section depends strongly on temperature and doping concentration ${\bf n_d}$. From experience we know that the effective resistance of the heater element is significantly lower than expected from the heater dimensions and the doping level. We attribute this observation to dopant diffusion during the high-temperature annealing step for activating the electrons in the high doped regions. Assuming a diffusivity of $D \, \simeq \, 10^{-12}$ cm$^2$s$^{-1}$ \cite{chang1963concentration,thai1970concentration} for the phosphorous diffusion at 1150$^\circ$ C and an annealing time of $t_a \, = \, 1800 \, {\rm s}$ we estimate that the doping concentration markedly increases from the nominal value of $3.3 \times 10^{17} \, {\rm cm}^{-3}$ at a distance of $\simeq 1$ $\mu$m from the heater ends. To account for this effect we introduce a continuous doping profile which is approximated by the known solution for the 1-dimensional step profile diffusion
\begin{equation}
n_{ld}(x) = {\bf n_d}+ (n_{hd}-{\bf n_d}) \left ( 1+ \frac{{\rm erf}\left ( \frac{x-l_h/2}{{\bf \Delta}} \right ) + {\rm erf} \left ( \frac{-x-l_h/2}{{\bf \Delta}} \right )}{2} \right )
\label{equdp}
\end{equation}
where $n_{hd} \, = \, 2.2 \times 10^{20} \, {\rm cm}^{-3}$ is the doping concentration from ion implantation in the low resistivity cantilever structure and $l_h \, = \, 4 \mu{\rm m}$ denotes the geometrical heater length defined by the manufacturing process and the x-coordinate points along the heater axis. Since neither the nominal doping concentration in the heater, $\bf n_d$, nor the dopant diffusivity expressed in terms of the parameter ${\bf \Delta} \, = \, 2 \sqrt{D t_a}$ are known with sufficient accuracy we treat these numbers as fit parameters. The local electrical conductance in the heater section is calculated following Ref. \cite{durig2005fundamentals}
\begin{equation}
\sigma_{ld} (x,T) = 1.6 \times 10^{-19} \, {\rm C} \left ( n_e(x,T) \mu_e(x,T) \, + \, n_h(x,T) \mu_h(x,T) \right )
\label{equ5}
\end{equation}
where the electron and hole densities are given by
\begin{eqnarray}
n_e(x,T) & = & \frac{1}{2} \left ( n_{ld}(x) + \sqrt{n_{ld}^2(x) + 4 n_i^2 (x,T)} \right )
\label{equ6} \\
n_h(x,T) & = & n_i^2(x,T) / n_e(x,T) \nonumber
\end{eqnarray}
with
\begin{eqnarray}
n_i(x,T) & \simeq & 2.70 \times 10^{19} \, {\rm cm^{-3}} \left (\frac{T}{300 \, {\rm K}} \right )^{3/2} \exp\left ( -\frac{E_g(x,T)}{2 k_B T} \right ) \,\,\,\, {\rm and}
\label{equ7} \\
E_g(x,T) & \simeq & 1.17 \, {\rm eV} - \frac{4.73 \times 10^{-4} \, {\rm eV \, K^{-1}} \, T^2}{T+ 636 \, {\rm K}}  - 0.025 \, {\rm eV} \times (n_{ld}(x)/10^{18} \, {\rm cm}^{-3})^{(1/3)} \nonumber
\end{eqnarray}
where the pre-factor in the $n_i$ expression is adjusted to reproduce the published value of $n_i(300\, {\rm K}) \, = \, 9.65 \times 10^9$ cm$^{-3}$ \cite{altermatt1999influence} in low doped silicon. The last term in the $E_g$-equation accounts for the so-called band gap narrowing due to the dopant atoms (see Figure 7 in Ref. \cite{berggren1981band}). The electron (effective mass $m_e \, \simeq \, 0.33$) and hole (effective mass $m_h \, \simeq \, 0.55$) mobilities can be expressed in terms of impurity scattering (denoted by the index $i$) and phonon scattering (denoted by the index $ph$) contributions
\begin{eqnarray}
\mu_e(x,T) & = & \frac{1}{\frac{1}{\mu_{i,e}(x)} + \frac{1}{\mu_{ph,e}(T)}}
\label{equ8} \\
\mu_h(x,T) & = & \frac{1}{\frac{1}{\mu_{i,h}}(x) + \frac{1}{\mu_{ph,h}(T)}}  \nonumber
\end{eqnarray}
with
\begin{eqnarray}
\mu_{i,e} (x) & \simeq & 4.1 \times 10^{11} \, {\rm cm^2 V^{-1} s^{-1}} \, \frac{1 {\rm cm^{-3/2}}}{n_{ld}^{1/2}(x)}
\label{equ9} \\
\mu_{i,h} (x) & = & \left ( \frac{m_e}{m_h} \right )^{1/2} \mu_{i,e}(x) \, \simeq \, 0.746 \, \mu_{i,e}(x) \nonumber
\end{eqnarray}
and
\begin{eqnarray}
\mu_{ph,e} (T) & \simeq & 1.6 \times 10^3 \, {\rm cm^2 V^{-1} s^{-1}} \left ( \frac {300 \, {\rm K}}{T} \right )^{\bf p}
\label{equ10} \\
\mu_{ph,h}(T) & = & \left ( \frac{m_e}{m_h} \right )^{5/2} \mu_{ph,e}(T) \, \simeq \, 0.279 \, \mu_{ph,e}(T) \,\,\, . \nonumber
\end{eqnarray}
The exponent ${\bf p}$ in the phonon mobility expression Eqn(\ref{equ10}) constitutes a fourth empirical fit parameter which is on the order of 2 as noted in Ref. \cite{durig2005fundamentals} whereas theoretical models would predict a value of 1.5.

Finally we follow Ref. \cite{chapman1963electrical} for parameterizing the temperature dependent conductivity $\sigma_{hd}(T)$ in the quasi-metallic high doped regions. It has been observed that $\sigma_{hd}^{-1}(T)$ can be described by a simple linear law for temperatures above 300 K and at doping levels greater than $\simeq \, 10^{19}$ cm$^{-3}$
\begin{equation}
\frac{1}{\sigma_{hd}(T)} = {\bf a_0} + {\bf a_1} \, (T-300 \, {\rm K}) \,\,\, .
\label{equ11}
\end{equation}
The coefficients ${\bf a_0}$ and ${\bf a_1}$ in the linear equation, which at a doping concentration of $1.3 \times 10^{20}$ cm$^{-3}$  are on the order of $6 \times 10^{-4}$ ohm cm and $1.6 \times 10^{-6}$ ohm cm K$^-1$, respectively (see Fig. 1 in Ref. \cite{chapman1963electrical}), constitute the last two fit parameters in the model.

\section{Raman temperature measurements}

To assess the simulation in terms of correctly predicting the tip temperature we performed Raman spectroscopy studies in analogy to the approach published in Refs \cite{abel2007thermal,nelson2007temperature}. The method exploits the temperature dependence of the resonance frequency of the T2g vibrational mode which is manifested by a shift of the corresponding Stokes peak in the Si Raman spectrum. The measurements were performed using a grating spectrometer (Horiba Jobin Yvon, LabRam HR) operated at 0.1 cm$^{-1}$ grating resolution and a Linkam TMS600 sample chamber which was adapted for vacuum ($\simeq 10^{-2}$ mbar) operation. The instrument is located in a sophisticated temperature and vibration controlled laboratory environment  \cite{lortscher2013next} to ensure that Raman measurements are not compromised by instrumental drift.

\begin{figure}[!] \centering
\includegraphics[width=60mm]{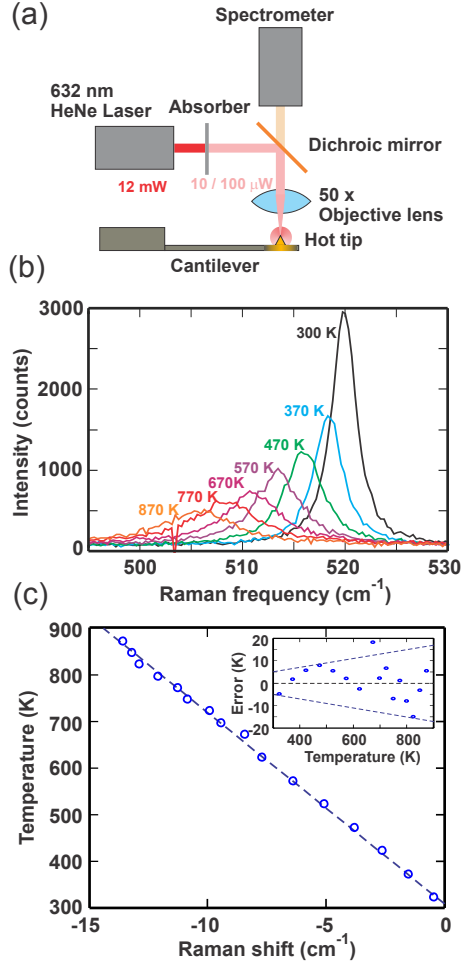}\\
\caption{\footnotesize (a) Schematic of the Raman measurement set-up. (b) Raman spectra recorded in the temperature range from 300 K to 870 K on a Si test sample. Note that the laser power (1 mW) and the integration time (30 s) have been adapted to yield a similar photon count as observed in the measurements of the tip temperature. (c) Raman shift versus sample temperature: Open circles correspond to measured data and the dashed line is a linear fit which constitutes the calibration function. Inset figure shows the deviation of the measure data from the calibration curve. The dashed lines indicate the calibration error which increases approximately linearly with temperature.} \label{figure-2}
\end{figure}

Illumination of the sample and collection of the scattered light is accomplished by means of a long working distance 50x objective lens with a numerical aperture of 0.5. The primary light is provided by a 632 nm HeNe laser source with an output power of 12 mW and using calibrated optical attenuators to reduce the beam power incident on the cantilever structure, see Fig. \ref{figure-2}(a). Specifically, Raman spectra were recorded at an incident power of 10 $\mu$W  at temperatures below 950 K and the power was increased to 100 $\mu$W above 950 K to compensate for the loss of signal strength with increasing temperature, see Fig. \ref{figure-2}(b). Considering the fact that approximately 10 \% of the laser light is actually absorbed in the cantilever structure the laser induced heating is not significant in comparison to the overall temperature error inferred from calibration experiments. The Raman spectra are recorded by placing the tip apex at the center of the objective focus. The Raman signal emanating from the tip is significantly enhanced owing to the field magnification effect at the tip apex which has a radius of curvature of less than 10 nm. Specifically we observed a 30x - 80x stronger Raman signal with respect to a flat Si surface. This enabled us to record clean spectra of the T2g optical phonon Stokes peak at $\simeq \, 520$ cm$^{-1}$ within 10 to 60 seconds depending on the tip temperature.

The temperature is inferred from measuring the shift of approximately -0.022 cm$^{-1}$K$^{-1}$ \cite{PhysRevB.1.638} of the T2g phonon frequency induced by thermal strain. The latter could also be influenced by mechanical stress in the structure leading to a shift on the order of $10^{-3}$ cm$^{-1}$MPa$^{-1}$ \cite{ganesan1970lattice}. From thermo-mechanical simulations we know that the stress levels in the heater section are below 20 MPa and consequently the stress induced error of the temperature measurement is less than 1 K and thus negligible. Calibration experiments were performed on a Si (100) wafer using the exact same set-up as for the cantilever measurement. The laser power was increased to 1 mW to compensate for the lower signal strength on planar samples. The Si wafer is uniformly heated using the hot plate in the sample stage from room temperature up to the maximum temperature of the hot plate of 900 K. Raman spectra are recorded at approximately 25 K temperature intervals and the signal is integrated over 30 seconds yielding spectra with approximately the same signal to noise ratio as in the cantilever measurements, see Fig. \ref{figure-2}(b). The Raman shifts $\Delta_R$ are obtained by fitting a Lorentzian profile to the measured spectral peaks. The temperature difference with respect to room temperature $T_{RT}$ can be fitted to a linear calibration function $T_{cal} - T_{RT} \, = \, 5\, {\rm K} \, - \, 41.1 \, {\rm K \, cm} \times \Delta_R$ which yields approximately a symmetric distribution of the calibration error $T_{cal}-T_{hot\,plate}$, see Fig. \ref{figure-2}(c). The calibration error can be parameterized in the form $\pm \delta T_{cal} \, \simeq \, 5 \, {\rm K} + 0.02 \times (T_{cal}-T_{RT})$. The constant 5 K term is a systematic error due to the fact that $T(\Delta_R)$ is not exactly a linear function over the full temperature range. Superimposed on this error is a seemingly random error due to an overall measurement inaccuracy which increases linearly with temperature.

\section{Comparison of modeling results with experimental data in the absence of coupling to air}

The observables of the measurement are the current flowing through the tip heater, $I_h$, the value of the electrical potential sensed by the right lever arm, $V_r$, and the tip temperature $T_{tip}$ while $V_b$ acts as control parameter, see Fig. \ref{figure-1} (b,c). $V_r$ measures the voltage drop occurring solely in highly doped Si and thus it helps to improve the fidelity of the simulation by decoupling the electrical conduction in low and high doped regions. The other important element is the series resistor, $R_s$, which ensures that $I_h (V_b)$ is a single valued function despite of the negative differential resistance exhibited by the tip heater when the intrinsic carrier density exceeds the doping concentration at high temperatures (typically above 800 K) \cite{chui1999intrinsic,durig2005fundamentals}.

The heat flux perpendicular to the lever plane is negligible in vacuum since there is no coupling to air and the black body radiation, on the order of 1 $\mu$W at the highest temperature of 1150 K, is more than three orders of magnitude smaller than the overall dissipated power in the lever ($\simeq$ 4 mW) at such heater temperatures. The thermal and electrical transport parameters in Eqs(\ref{equ4}-\ref{equ11}) are assumed to be constants across the lever cross section. Moreover, the lever thickness is constant over the entire structure. Therefore, a homogeneous temperature and current distribution along the out of plane axis is obtained and the simulation problem is reduced to a 2-dimensional one. We use the the FEniCS finite element solver \cite{logg2012automated,alnaes2015fenics} to calculate the temperature and voltage distributions in the lever for a given set of fit parameters and a range of bias potentials $V_b$ from which we then compute the observables $I_h (V_b)$, $V_r(V_b)$, and $T_{tip}(V_b)$. The computed data serves as input for a non-linear Levenberg-Marquardt least square fit implemented in Python using the LMFIT function \cite{lmfit} to find the optimal parameter set which best fits the corresponding measured data.

\begin{figure}[!] \centering
\includegraphics[width=100mm]{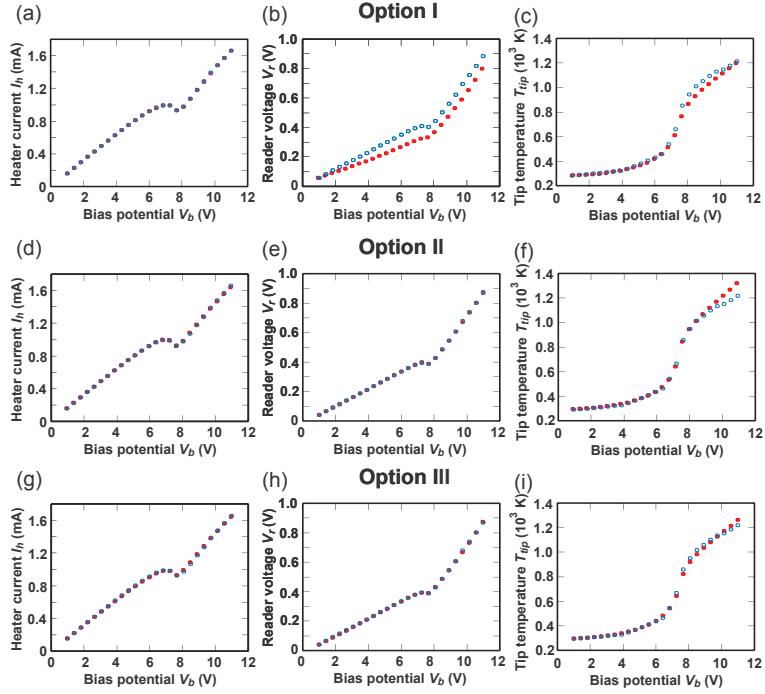}\\
\caption{\footnotesize (a,d,g) Heater current $I_h$, (b,e,g) reader voltage $V_r$, and (c,f,i) tip temperature $T_{tip}$ as a function of bias potential. Open circles correspond to the measured data and full circles correspond to the model prediction. Option I (panels (a)-(c)), II (panels (d)-(f)), and III (panels ((g)-(i)) respectively refer to the input vector $\{I_h(V_b)\}$, $\{I_h(V_b), V_r(V_b)\}$, and $\{I_h(V_b),V_r(V_b),T_{tip}(V_b)\}$ used in the parameter optimization.} \label{figure-3}
\end{figure}

\begin{figure}[!] \centering
\includegraphics[width=100mm]{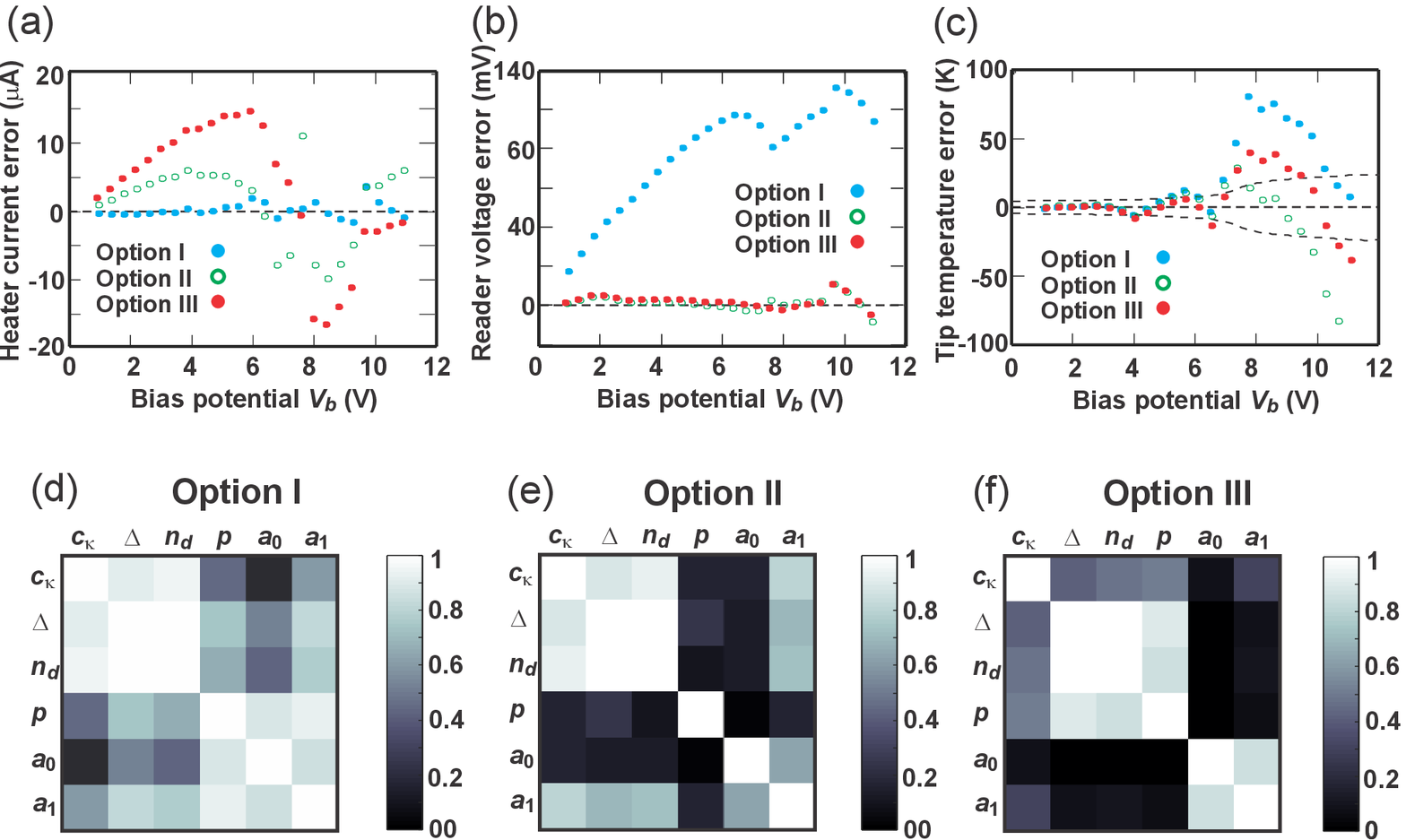}\\
\caption{\footnotesize Error of the predicted observables versus bias voltage $V_b$ for the three fit options I-III (measured values minus predicted values): (a) Heater current $I_h$, (b) reader voltage $V_r$, (c) tip temperature $T_{tip}$. The dashed lines in (c) indicate the expected calibration error of the Raman temperature measurement. (d) - (f) Covariance matrix of the parameter fit normalized by the diagonal elements for option I-III. Fit parameters are: Scale parameter for the thermal conductivity ${\bf c_\kappa}$ (see Eqn(\ref{equ4})), width of the doping profile ${\bf \Delta}$ and nominal doping level ${\bf n_d}$ of the heater (see Eqn(\ref{equdp})), phonon scattering exponent ${\bf p}$ (see Eqn(\ref{equ10})), and electrical resistivity in highly doped Si expressed in terms of the parameters ${\bf a_0}$ and ${\bf a_1}$ (see Eqn(\ref{equ11})). Option I - III respectively refer to the input vector $\{I_h(V_b)\}$, $\{I_h(V_b), V_r(V_b)\}$, and $\{I_h(V_b),V_r(V_b),T_{tip}(V_b)\}$ used in the parameter optimization.} \label{figure-4}
\end{figure}

The measurements of $I_h(V_b)$, $V_r$, and $T_{tip}(V_b)$ are performed simultaneously in the evacuated sample chamber of the Raman set-up. $V_b$ is ramped from 1 V to 11 V in 25 equally spaced increments and the cantilever is allowed to equilibrate for 10 seconds before data is taken. Three types of simulations termed optimization I, II, and III are performed using $\{I_h(V_b)\}$, $\{I_h(V_b), V_r(V_b)\}$, and $\{I_h(V_b),V_r(V_b),T_{tip}(V_b)\}$ as input vector in the parameter optimization, respectively. Equal weight is assigned to the observables in the fit procedure by normalizing the observables by their respective maximum values. Approximately 100 iterations are needed to obtain a set of fit parameters with less than $10^{-5}$ relative error with respect to the true optimum. Each iteration involves 7 FEniCS simulations to compute the derivative matrix for the least square fit program each of which consumes about 2 minutes of computation time on a workstation computer.

\begin{figure}[!] \centering
\includegraphics[width=60mm]{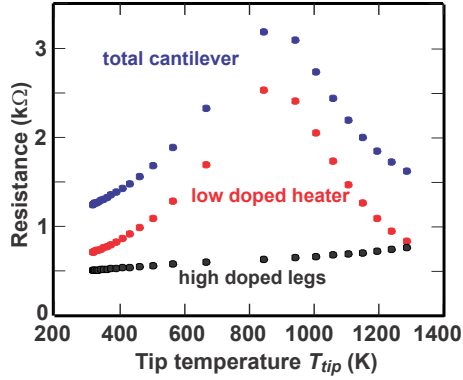}\\
\caption{\footnotesize Resistance of the total cantilever, the low doped heater element, and the high doped leg structure versus heater temperature as determined from the simulation.} \label{figure-5}
\end{figure}

A comparison of the measured and simulated data are shown in Figs \ref{figure-3} and \ref{figure-4}. Using solely $I_h(V_b)$ as input vector (optimization I) results in a significant error in the prediction of the reader potential, $V_r$, and tip temperature, $T_{tip}$. As can bee seen from the correlation matrix, Fig. \ref{figure-4}(d), all fit parameters are strongly interdependent among each other making the parameter estimation prone to systematic errors due to model inadequacies. Adding $V_r(V_b)$ to the input vector (optimization II) removes the $V_r$ error and a significant improvement of the model accuracy is obtained for the tip temperature up to 1100 K. One also sees from the correlation matrix, see Fig. \ref{figure-4}(e), that a much better decoupling of the fit parameters is obtained. Including the tip temperature data to the fit vector (option III) yields excellent agreement between predicted and measured tip temperature over the entire temperature range. However, the current error increases somewhat but the relative error remains below 2 \%. More importantly, we achieve an excellent mutual decoupling between the thermal conductivity parameter ${\bf c_\kappa}$, the electrical leg resistance parameters ${\bf a_0 }$ and ${\bf a_1}$, and the electrical heater parameters ${\bf \Delta , \, n_d }$, and ${\bf p}$, see Fig. \ref{figure-4}(f). The values of the fit parameters obtained from the three otpimization schemes are listed in Table \ref{table-1}. It is apparent that the most sensitive parameters are the doping level of the heater, ${\bf n_d}$ and the phonon scattering exponent ${\bf p}$ which are significantly different depending on the optimization model. On the other hand, robust predictions for ${\bf c_\kappa, \, \Delta, \, a_0}$, and ${\bf a_1}$ parameters are obtained in particular if one compares optimization II and III. Given the model parameters one can easily compute the resistances of the heater, $R_h$, the leg structure, $R_{l1}+R_{l2}$, and the total cantilever, $R_h+R_{l1}+R_{l2}$, as a function of tip temperature, see Fig. \ref{figure-5}. As expected, the overall resistance of the cantilever is dominated by the heater resistance up to a heater temperature of 1200 K and only at temperatures larger than 1300 K the leg resistance starts to take over. Nevertheless, we point out that one needs to account for the temperature dependence of the leg resistance in order to achieve good agreement of the modeling results with the measured data.

\begin{table} \centering
\normalsize{
\begin{tabular}{|c|c|c|c|c|c|c|c|c|c|} \hline
$\mathbf{optimization}$ & $\mathbf{c_\kappa}$ & $\mathbf{\Delta}$ & $\mathbf{n_d}$ & $\mathbf{p}$ & $\mathbf{a_0}$ & $\mathbf{a_1}$ & $\mathbf{T_i}$ \\
$\mathbf{vector}$ & & $\mathbf{\mu m}$ & $\mathbf{10^{17} \, cm^{-3}}$ & & $\mathbf{10^{-4}\, ohm \, cm}$ & $\mathbf{10^{-6} \, ohm \, cm \, K^{-1}}$ & $\mathbf K$\\
\hline
$\mathbf{I_h}$ & 0.744(7) & 0.681(30) & 4.10(62) & 2.64(2) & 5.87(21) & 1.67(10) & 805  \\
$\mathbf{I_h, \, V_r}$ & 0.675(8) & 0.649(53) & 6.87(140) & 2.80(2) & 9.32(6) & 1.08(3) & 851  \\
$\mathbf{I_h, \, V_r, \, T_{tip}}$ & 0.686(5) & 0.647(51) & 5.75(94) & 2.65(5) & 9.34(15) & 1.11(5) & 831 \\
\hline
\end{tabular}}
\caption{\footnotesize Compilation of parameter values obtained form fitting the simulation data to the experimental data using the observables $I_h(V_b)$ (option I), $I_h(V_b)\, - \, V_r(V_b)$ (option II), and $I_h(V_b) \, - \, V_r(V_b) \, - \, T_{tip}(V_b)$ (option III) as input vector for the fit. The numbers in brackets refer to the fit uncertainty in least significant figures. The inversion temperature $T_i$ corresponding to the doping concentration ${\bf n_d}$ and ${\bf p}$ value is tabulated in the last column.} \label{table-1}
\end{table}

\begin{table} \centering
\normalsize{
\begin{tabular}{|c|c|c|c|c|c|c|c|c|} \hline
$\mathbf{optimization \, III}$ & $\mathbf{c_\kappa}$ & $\mathbf{\Delta}$ & $\mathbf{n_d}$ & $\mathbf{p}$ & $\mathbf{a_0}$ & $\mathbf{a_1}$  \\
$\mathbf{I_h, \, V_r, \, T_{tip}}$ & & $\mathbf{\mu m}$ & $\mathbf{10^{17} \, cm^{-3}}$ & & $\mathbf{10^{-4}\, ohm \, cm}$ & $\mathbf{10^{-6} \, ohm \, cm \, K^{-1}}$ \\
\hline
$\mathbf{Lever \, 1}$ & 0.686 & 0.647 & 5.75 & 2.65 & 9.34 & 1.11  \\
$\mathbf{n_d \, fixed}$ & 0.692 & 0.763 & $\mathbf{3.30}$ & 2.74 & 9.30 & 1.12  \\
\hline
$\mathbf{Lever \, 2}$ & 0.687 & 0.719 & 3.64 & 2.80 & 10.2 & 1.25 \\
$\mathbf{n_d \, fixed}$ & 0.688 & 0.735 & $\mathbf{3.30}$ & 2.81 & 10.2 & 1.26  \\
\hline
$\mathbf{Lever \, 3}$ & 0.688 & 0.646 & 3.50 & 2.85 & 10.3 & 1.26 \\
$\mathbf{n_d \, fixed}$ & 0.690 & 0.659 & $\mathbf{3.30}$ & 2.86 & 10.3 & 1.26  \\
\hline
$\mathbf{Lever \, 4}$ & 0.613 & 0.544 & 5.93 & 2.86 & 10.7 & 1.23 \\
$\mathbf{n_d \, fixed}$ & 0.686 & 0.716 & $\mathbf{3.30}$ & 2.99 & 10.7 & 1.26  \\
\hline
\end{tabular}}
\caption{\footnotesize Compilation of parameter values obtained form fitting the simulation data obtained from four different levers to the experimental data using optimization vector $I_h(V_b) \, - \, V_r(V_b) \, - \, T_{tip}(V_b)$ as input vector for the fit (option III). The results discussed in the manuscript refer to lever 1. The parameter values obtained by fixing the doping concentration to the nominal value of $n_d \, = \, 3.3 \times 10^{17}$ cm$^{-3}$ are also quoted for each lever.} \label{table-2}
\end{table}

To assess the robustness of the model we performed measurements and model fitting on three additional cantilevers taken from the same wafer batch. In all cases we obtain excellent consistency of the model parameters within less than 10 \% error margin with the exception of the doping concentration of the heater which exhibits a significantly larger scatter, see Table \ref{table-2}. We attribute this parameter sensitivity to the short length of the heater. We also ran model fits fixing the doping concentration in eqn(\ref{equdp}) to the nominal value of ${\bf n_d} \, = \, 3.3 \times 10^{17}$ cm$^{-3}$ which has little effect on the other parameters. Therefore we can say that the model not only reliably reproduces the electrical characteristics of our cantilever design but one also obtains accurate measurements of the physical parameters used to describe the thermal and electrical conduction in the high and low doped silicon. We also note that the model provides reliable numbers for the dopant diffusion at the interface between high doped leg structures to the low doped heater section in a non destructive experiment. This property is highly valuable for the control and optimization of the critical thermal annealing process used in the fabrication of the cantilevers.

\begin{figure}[!] \centering
\includegraphics[width=100mm]{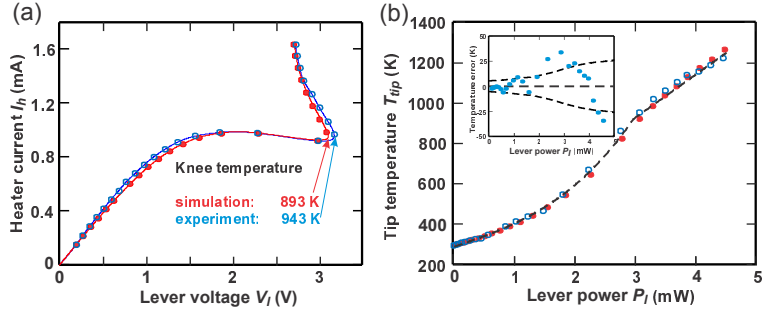}\\
\caption{\footnotesize (a) Heater current $I_h$ versus lever potential $V_l$: Open circles correspond to measured data and filled circles correspond to the model prediction using parameters from the option III fit. Note that the simulation predicts a knee temperature (indicated by the arrows) which is 50 K lower than the measured value. (b) Tip temperature $T_{tip}$ versus electrical power $P_l \, = \, V_l \times I_h$ dissipated in the lever: Open circles correspond to measured data and filled circles correspond to the model prediction using parameters from the option III fit. The dashed line represents the algebraic form (see Eqn(\ref{equ12})) of the temperature-power correlation. The inset panel shows the difference between measured and predicted tip temperature as a function of $P_l$ (The dashed lines indicate the expected calibration error of the Raman temperature measurement.).} \label{figure-6}
\end{figure}

\begin{figure}[!] \centering
\includegraphics[width=60mm]{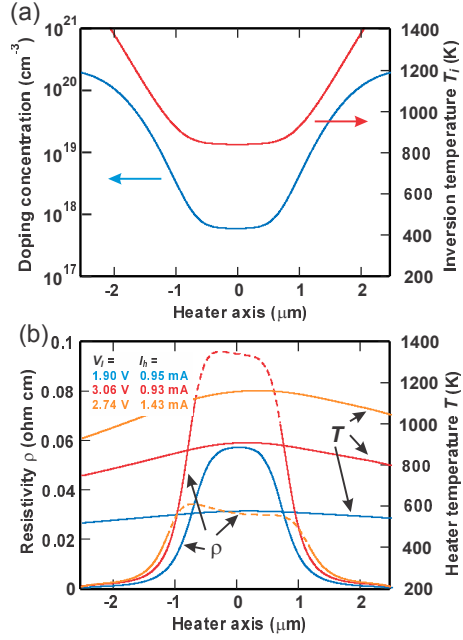}\\
\caption{\footnotesize (a) Doping profile calculated according to Eqn(\ref{equdp}) and inversion temperature in the heater section for ${\bf \Delta}$ = 0.647 $m$m and ${\bf n_d}$ = 5.75$\times 10^{17}$ cm$^{-3}$, see option III in Table \ref{table-1}. The x-axis is centered with respect to the heater which has a nominal length of 4 $\mu$m. Note that the effective heater length is reduced by approximately a factor of 2 due the diffusion of doping atoms from the highly doped connecting Si structures during the thermal activation and annealing step in the fabrication process. (b) Local resistivity $\rho$ and local heater temperature $T$ as determined from the simulation at different operating points. Solid lines in the $\rho$-curves correspond to a positive value of $d \rho / d T$ whereas negative values are obtained in the dashes sections. The curve labeled ($V_l$ = 3.06 V / $I_h$ = 0.93 mA) corresponds to the knee point at $T_{tip}$ = 893 K indicated in Fig. \ref{figure-5}. The curves labeled ($V_l$ = 1.90 V / $I_h$ = 0.95 mA) and ($V_l$ = 2.74 V / $I_h$ = 1.43 mA) represent situations in the plateau region below the knee point and in the high temperature region above the knee point, respectively.} \label{figure-7}
\end{figure}

The more standard representation of the electrical lever characteristics is in the form of a so-called I-V curve which is a map of the heater current $I_h$ observed at a lever potential $V_l \, = \, V_b - R_s I_h$, see Fig. \ref{figure-6}(a). The prominent feature in the I-V curve is the knee-point at which the differential lever resistance assumes a zero value, i.e. $d V_l / d I_h \, = 0$. This condition is equivalent to $d R_l / dT \, = \, 0$ where $R_l \, = \, V_l/I_h$ denotes the lever resistance. The knee point has often been used as reference point for the temperature calibration assuming that the temperature at this point coincides with the inversion temperature $T_i$ of the heater. The latter is defined by $d \rho (T_i) / dT \, = \, 0$. From the heater doping level of ${\bf n_d} \, = \, 5.75 \times 10^{17}$ cm$^{-3}$ we obtain $T_i \, = \, 831$ K, see Table \ref{table-1}, which is significantly lower than the predicted, 893 K, and the measured, 943 K, values. There are two reasons for this discrepancy: (1) The resistivity of the high doped Si increases with temperature which needs to be compensated by a corresponding negative temperature coefficient of the resistance of the heater section. This effect accounts for an approximately 15 K shift of the knee temperature towards a higher value. (2) The doping profile in the heater is highly non-uniform, see Fig. \ref{figure-7}(a), and as a result, the inversion temperature is also non-uniform. To obtain an overall negative differential heater resistance a significantly higher temperature is required in order to compensate the positive differential resistance from the parts of the heater in which the temperature is below the local inversion temperature, see Fig. \ref{figure-7}(b). Another consequence of the non-uniform doping profile is the fact that the effective heater width increases with temperature and as a result the heater becomes more efficient with increasing temperature.

Despite the rather sophisticated model used in the simulation we have difficulties to accurately predict the I-V curve in the vicinity of the knee point. In particular, the predicted knee temperature is systematically underestimated by $\simeq$ 50 K. This observation tells us that the I-V curve is extremely sensitive to how accurately one models the heater section. Ignoring the tip in the model might be one of the reasons for this shortcoming. We also did not include the Joule-Thomson effect in the model. However, we know from other simulations that the Joule-Thomson effect does not fundamentally improve the prediction accuracy but it leads to a shift of the temperature distribution by less than 500 nm in the direction of the current flow.

More interesting from a temperature calibration point of view is the fact that an excellent agreement between measured and predicted tip temperature is obtained if the temperature is plotted as a function of the electrical power $P_l \, = \, V_l I_h$ dissipated in the lever, see Fig. \ref{figure-6}(b). The correlation can be expressed in terms of a simple algebraic equation yielding less than 25 K temperature error over the entire temperature range
\begin{eqnarray}
T_{tip} & = & T_{RT} \, + \, 103 \, {\rm K} \times \frac{P_l}{1 \, {\rm mW}} \, + \, 11.2 \, {\rm K } \times \left ( \frac{P_l}{1 \, {\rm mW}} \right )^{3.1}
\label{equ12} \\
& {\rm for} & P_l \, \le \, P_{th} \, = \, 2.95 \, {\rm mW} \, \equiv \, T_{tip} \, \le \, T_{th} \, = \, 916 \, {\rm K} \nonumber \\
{\rm and} & & \nonumber \\
T_{tip} & = &  T_{RT} \, + \, 213 \, {\rm K} \times \frac{P_l}{1 \, {\rm mW}} \nonumber \\
& {\rm for} & P_l \, > \, P_{th} \, \equiv \, T_{tip} \, > \, T_{th} \nonumber
\end{eqnarray}
where $T_{RT}$ denotes the tip temperature at zero power. The nice feature of the power curve is the sharp transition from non-linear to linear behavior at the threshold temperature $T_{th}$ which provides a good estimate for the actual knee temperature. The non-linear dependence of the tip temperature as a function of applied power below $T_{th}$ is mainly caused by the decrease of the thermal conductivity in Si with increasing temperature which helps to confine the heat within the heater. Note, however, that the fit exponent of 3.1 expresses merely an observed correlation and its value is neither universal nor does it have a deeper physical meaning.  The linear relation above $T_{th}$ is a robust feature which has been observed in simulations of other lever structures and doping levels in the heater. It appears to be a rather fortuitous coincidence which is hard to explain in simple terms.

\section{Cantilever modeling in the presence of air}

\begin{figure}[!] \centering
\includegraphics[width=100mm]{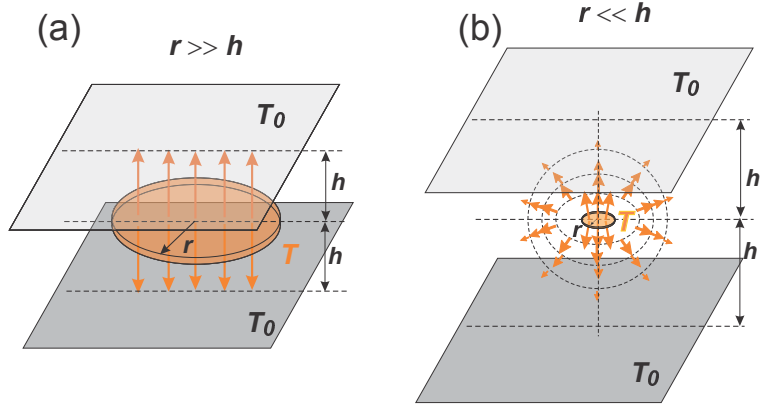}\\
\caption{\footnotesize Schematic of the heat flux emanating from a circular disk heater with radius $r$ and at a temperature $T$. The heater is placed midway between two planar surfaces which are at a temperature $T_0$ and which are separated by a distance $2 h$. (a) If the radius of the heater is larger than the distance from the planes, the heat flux can be approximated by the constant temperature gradient solution and correspondingly the flux density scales as $1/h$. (b) In the opposite limit, $r < h$, the heat flux is governed by the $1/d$ scaling of temperature gradient where $d$ denotes the distance from the heater leading to a flux funneling towards the heater. As a result, the mean heat flux density scales as $1/r$.} \label{figure-8}
\end{figure}

Including the thermal coupling to air substantially increases the complexity of the modeling task because the heat flux per unit area strongly depends on the geometry. To illustrate this property we consider a circular disk of radius $r$ which is at a temperature $T$, see Fig. \ref{figure-8}. The disk is positioned midway between two infinite planes at a temperature $T_0$ and the planes are separated by a distance $2 h$. If $h \ll r$ the heat flux is given by $j \, = \, \kappa_{air} (T-T_0)/h$. Correspondingly the total heat current $J \, = \, 2 \pi r^2 j$ scales as the area of the disk and inversely with $h$. Alternatively, one can describe the heat loss due to the thermal conduction in air in terms of a transfer coefficient $K_{air} \, = \, J/((T-T_0)2 \pi r^2) \, = \, \kappa_{air}/h \, \simeq \, 2.4 \times 10^4 \, {\rm Wm}^{-2} {\rm K}^{-1} \times (1 \, {\rm \mu m}/h)$. In the other extreme, $h \gg r$, the heat current emanating from the disk is described by the spreading resistance expression $J \, = \, 4 \kappa_{air} r \, (T-T_0)$ \cite{yovanovich1970temperature} which scales proportional to the radius, and we obtain for the transfer coefficient $K_{air} \, = \, J/((T-T_0)2 \pi r^2) \, = \, (2/\pi)\kappa_{air}/r \, \simeq \, 1.5 \times 10^4 \, {\rm Wm}^{-2} {\rm K}^{-1} \times (1 \, {\rm \mu m}/r)$. The coupling coefficient depends on the size of the structure and it can assume values on the order of $10^4 \, {\rm Wm}^{-2} {\rm K}^{-1}$ at the micrometer scale. In other words, the heat flux into air is governed either by the distance to the nearest heat sink or the dimension of the heated structure depending upon whichever is the smallest. Therefore one cannot simply use some global coupling resistance as has been done in some of the published work.

\begin{figure}[!] \centering
\includegraphics[width=100mm]{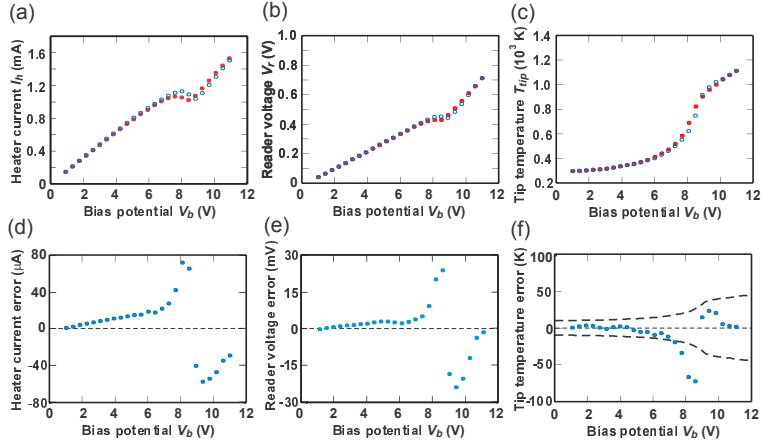}\\
\caption{\footnotesize (a) Heater current $I_h$, (b) reader voltage $V_r$, and (c) tip temperature $T_{tip}$ as a function of bias potential for a lever structure operated in air. Open circles correspond to the measured data and full circles correspond to the model prediction using the parameter set determined from the vacuum simulation (option III) and the expression in Eqn(\ref{equ13}) for the thermal conductivity in air. (d) - (f) Error of the model prediction corresponding to the observables in (a) - (c). The dashed lines in (f) indicate the expected calibration error of the Raman temperature measurement.} \label{figure-9}
\end{figure}

The only viable way to treat the heat flux into the surrounding air in a physical meaning full way is to solve Eqs(\ref{equ1}-\ref{equ3}) in a full 3-D simulation which adds substantial complexity to the modeling task (typically 3 orders of magnitude more execution time for running a finite element simulation). Here we consider a cantilever which is freely suspended in air as an approximation to the real experimental situation in the measurement set-up. The air volume is meshed in a 60 $\mu$m radius sphere around the cantilever and the open boundary problem is solved using the Kelvin transform for the exterior space \cite{freeman1989open}. The boundary conditions for the temperature are $T \, = \, 293$ K at infinity and at the end sections of the lever. The thermal conductivity of air for the temperature range pertaining to our problem is parameterized as follows \cite{kadoya1985viscosity}
\begin{eqnarray}
\kappa_{air} & = & 1\, {\rm W cm^{-1}K^{-1}} \times \left (2.60 \times 10^{-4} + 4.70 \times 10^{-7} \, {\rm K^{-1}} \, T + 1.48 \times 10^{-7} \, {\rm K^{-1/2}} \, T^{1/2} - \right . \nonumber \\
 & & \left . - 6.63 \times 10^{-2} \, {\rm{K}} \, T^{-1} + 9.14 \, {\rm K^2} \, T^{-2} - 6.50 \times 10^2 \, {\rm K^3} \, T^{-3} \right ) \,\,\, .
\label{equ13}
\end{eqnarray}
To simplify matters and as a test for the predictive power of the model we use the material and heater parameters determined from optimization III of the vacuum data. The agreement between the predicted and measured observables is indeed very good, see Fig. \ref{figure-9} considering the fact that there are no adjustable parameters in the simulation.

\begin{figure}[!] \centering
\includegraphics[width=100mm]{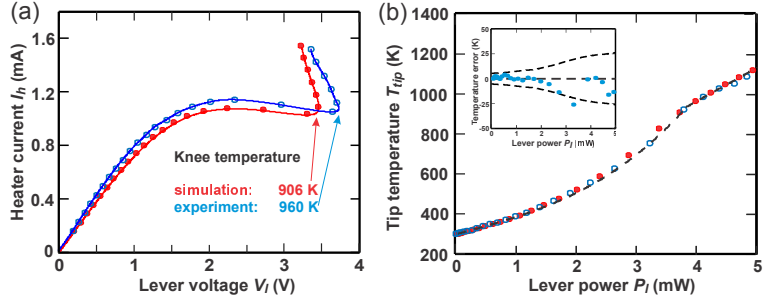}\\
\caption{\footnotesize (a) Heater current $I_h$ versus lever potential $V_l$ for a lever structure operated in air: Open circles correspond to measured data and filled circles correspond to the model prediction using parameters from the vacuum simulation (option III) and the expression in Eqn(\ref{equ13}) for the thermal conductivity in air. Note that the simulation predicts a knee temperature (indicated by the arrows) which is 54 K lower than the measured value. (b) Tip temperature $T_{tip}$ versus electrical power $P_l \, = \, V_l \times I_h$ dissipated in the lever: Open circles correspond to measured data and filled circles correspond to the model prediction. The dashed line represents the algebraic form of the temperature-power correlation substituting $P_l /k$ in Eqn(\ref{equ12}) with k=1.29. The inset panel shows the difference between measured and predicted tip temperature as a function of $P_l$ (The dashed lines indicate the expected calibration error of the Raman temperature measurement).} \label{figure-10}
\end{figure}

As expected from the discussion in the context of the vacuum results the predicted I-V curve deviates substantially from the measured one, see Fig. \ref{figure-10}(a), and the simulation again predicts a knee temperature which is a bit more than 50 K below the measured value. However, as was already the case before, the tip temperature versus power characteristics of the simulation and the measurement overlap with less than 25 K error. The intriguing finding of this study is the fact that one can use exactly the same functional form for the correlation function to represent the data by merely scaling $P^{vacuum}_l \, = \, P^{air}_l/k$ in Eqn(\ref{equ12}) by a constant factor $k \, = \, 1.29$. In other words, the additional heat loss due to the heat conduction in air is a constant fraction, namely 0.29, of the total power dissipated in the lever irrespective of the tip temperature. One may expect that this property is also approximately preserved in more complex situations involving for example air cooling by a nearby surface on one side of the lever and open space on the other side. The scale factor will be different, of course, but it can be determined from a measurement of the lever power $P_{knee}$ at the knee point and using the relation $P_{knee} \, = \, k \times P_{th}$ where $P_{th}$ is determined from a much simpler vacuum simulation.

\section{Conclusions}

One may argue that the excellent predictive power of our simulation is simply due to the large number of fit parameters used in the model. This view ignores the fact that the fit parameters have a distinct physical meaning and are not just expansion coefficients in an empirical correlation function. As such, our model is robust in the sense that the we obtain a consistent set of fit parameters from different cantilever specimens. This allows us to pin down material parameters with better than 10\% reliability, which is remarkable given the complex geometry and device physics, in particular of the low doped heater section. We also demonstrate that dopant diffusion induced by a thermal annealing step in the fabrication process is significant and we can actually measure the diffusion length in a non-destructive manner. We also find that the I-V curve is extremely hard to predict accurately at the knee point despite the fact that we use a rather sophisticated tapered doping profile model to describe the low-doped heater section. Ignoring the presence of the tip in the numerical model could be responsible for this shortcoming. This hypothesis is supported by the air simulations. Here the influence of the tip is further accentuated by the effective air coupling provided by the sharp apex structure which leads to an additional heat loss channel. Ignoring this channel could be the cause for the significant discrepancy between predicted and measured I-V curves in Fig. \ref{figure-9}(a). On the other hand, we find that the correlation between tip temperature and electrical power dissipated in the lever is accurately reproduced by our simulations both in vacuum and in air. A particularly intriguing result is the fact that the correlation in air and in vacuum can be expressed in terms of exactly the same analytical function by merely applying a linear scaling of the power. In other words, the heat which is globally dissipated into air is a constant fraction of the applied electrical power irrespective of the tip temperature. This result, which is hard to rationalize given the complex geometry dependence of the heat dissipation into air and the highly temperature dependent thermal conductivity in air, in combination with I-V curve measurements is of significant practical value for an accurate prediction of the tip temperature under various degrees of air cooling in real experimental situations.

\begin{acknowledgments}
We thank Ute Drechsler for the fabrication of the cantilevers.  The reported research received funding from the  European  Union's Seventh Framework Program FP7/2007-2013 under Grant Agreement No. 318804 (Single Nanometer Manufacturing for beyond CMOS Devices, acronym SNM) and through the European Research Council StG No. 307079.
\end{acknowledgments}


\begin{thebibliography}{43}
\expandafter\ifx\csname natexlab\endcsname\relax\def\natexlab#1{#1}\fi
\expandafter\ifx\csname bibnamefont\endcsname\relax
  \def\bibnamefont#1{#1}\fi
\expandafter\ifx\csname bibfnamefont\endcsname\relax
  \def\bibfnamefont#1{#1}\fi
\expandafter\ifx\csname citenamefont\endcsname\relax
  \def\citenamefont#1{#1}\fi
\expandafter\ifx\csname url\endcsname\relax
  \def\url#1{\texttt{#1}}\fi
\expandafter\ifx\csname urlprefix\endcsname\relax\def\urlprefix{URL }\fi
\providecommand{\bibinfo}[2]{#2}
\providecommand{\eprint}[2][]{\url{#2}}

\bibitem[{\citenamefont{Majumdar et~al.}(1993)\citenamefont{Majumdar, Carrejo,
  and Lai}}]{majumdar1993thermal}
\bibinfo{author}{\bibfnamefont{A.}~\bibnamefont{Majumdar}},
  \bibinfo{author}{\bibfnamefont{J.}~\bibnamefont{Carrejo}}, \bibnamefont{and}
  \bibinfo{author}{\bibfnamefont{J.}~\bibnamefont{Lai}},
  \bibinfo{journal}{Appl. Phys. Lett.} \textbf{\bibinfo{volume}{62}},
  \bibinfo{pages}{2501} (\bibinfo{year}{1993}).

\bibitem[{\citenamefont{Mamin and Rugar}(1992)}]{mamin1992thermomechanical}
\bibinfo{author}{\bibfnamefont{H.}~\bibnamefont{Mamin}} \bibnamefont{and}
  \bibinfo{author}{\bibfnamefont{D.}~\bibnamefont{Rugar}},
  \bibinfo{journal}{Appl. Phys. Lett.} \textbf{\bibinfo{volume}{61}},
  \bibinfo{pages}{1003} (\bibinfo{year}{1992}).

\bibitem[{\citenamefont{Chui et~al.}(1997)\citenamefont{Chui, Mamin, Terris,
  Rugar, Goodson, and Kenny}}]{chui1997micromachined}
\bibinfo{author}{\bibfnamefont{B.~W.} \bibnamefont{Chui}},
  \bibinfo{author}{\bibfnamefont{H.~J.} \bibnamefont{Mamin}},
  \bibinfo{author}{\bibfnamefont{B.}~\bibnamefont{Terris}},
  \bibinfo{author}{\bibfnamefont{D.}~\bibnamefont{Rugar}},
  \bibinfo{author}{\bibfnamefont{K.~E.} \bibnamefont{Goodson}},
  \bibnamefont{and} \bibinfo{author}{\bibfnamefont{T.~W.} \bibnamefont{Kenny}},
  in \emph{\bibinfo{booktitle}{Solid State Sensors and Actuators, 1997.
  TRANSDUCERS'97 Chicago., 1997 International Conference on}}
  (\bibinfo{organization}{IEEE}, \bibinfo{year}{1997}),
  vol.~\bibinfo{volume}{2}, pp. \bibinfo{pages}{1085--1088}.

\bibitem[{\citenamefont{Chui et~al.}(1996)\citenamefont{Chui, Stowe, Kenny,
  Mamin, Terris, and Rugar}}]{chui1996low}
\bibinfo{author}{\bibfnamefont{B.}~\bibnamefont{Chui}},
  \bibinfo{author}{\bibfnamefont{T.}~\bibnamefont{Stowe}},
  \bibinfo{author}{\bibfnamefont{T.}~\bibnamefont{Kenny}},
  \bibinfo{author}{\bibfnamefont{H.}~\bibnamefont{Mamin}},
  \bibinfo{author}{\bibfnamefont{B.}~\bibnamefont{Terris}}, \bibnamefont{and}
  \bibinfo{author}{\bibfnamefont{D.}~\bibnamefont{Rugar}},
  \bibinfo{journal}{Appl. Phys. Lett.} \textbf{\bibinfo{volume}{69}},
  \bibinfo{pages}{2767} (\bibinfo{year}{1996}).

\bibitem[{\citenamefont{Vettiger et~al.}(2002)\citenamefont{Vettiger, Cross,
  Despont, Drechsler, Durig, Gotsmann, Haberle, Lantz, Rothuizen, Stutz
  et~al.}}]{vettiger2002millipede}
\bibinfo{author}{\bibfnamefont{P.}~\bibnamefont{Vettiger}},
  \bibinfo{author}{\bibfnamefont{G.}~\bibnamefont{Cross}},
  \bibinfo{author}{\bibfnamefont{M.}~\bibnamefont{Despont}},
  \bibinfo{author}{\bibfnamefont{U.}~\bibnamefont{Drechsler}},
  \bibinfo{author}{\bibfnamefont{U.}~\bibnamefont{Durig}},
  \bibinfo{author}{\bibfnamefont{B.}~\bibnamefont{Gotsmann}},
  \bibinfo{author}{\bibfnamefont{W.}~\bibnamefont{Haberle}},
  \bibinfo{author}{\bibfnamefont{M.}~\bibnamefont{Lantz}},
  \bibinfo{author}{\bibfnamefont{H.}~\bibnamefont{Rothuizen}},
  \bibinfo{author}{\bibfnamefont{R.}~\bibnamefont{Stutz}},
  \bibnamefont{et~al.}, \bibinfo{journal}{IEEE T. Nanotechnol.}
  \textbf{\bibinfo{volume}{1}}, \bibinfo{pages}{39} (\bibinfo{year}{2002}).

\bibitem[{\citenamefont{Gom{\`e}s et~al.}(2015)\citenamefont{Gom{\`e}s, Assy,
  and Chapuis}}]{gomes2015scanning}
\bibinfo{author}{\bibfnamefont{S.}~\bibnamefont{Gom{\`e}s}},
  \bibinfo{author}{\bibfnamefont{A.}~\bibnamefont{Assy}}, \bibnamefont{and}
  \bibinfo{author}{\bibfnamefont{P.-O.} \bibnamefont{Chapuis}},
  \bibinfo{journal}{Phys. Status Solidi A} \textbf{\bibinfo{volume}{212}},
  \bibinfo{pages}{477} (\bibinfo{year}{2015}).

\bibitem[{\citenamefont{Garcia et~al.}(2014)\citenamefont{Garcia, Knoll, and
  Riedo}}]{garcia2014advanced}
\bibinfo{author}{\bibfnamefont{R.}~\bibnamefont{Garcia}},
  \bibinfo{author}{\bibfnamefont{A.~W.} \bibnamefont{Knoll}}, \bibnamefont{and}
  \bibinfo{author}{\bibfnamefont{E.}~\bibnamefont{Riedo}},
  \bibinfo{journal}{Nat. Nanotechnol.} \textbf{\bibinfo{volume}{9}},
  \bibinfo{pages}{577} (\bibinfo{year}{2014}).

\bibitem[{\citenamefont{Szoszkiewicz et~al.}(2007)\citenamefont{Szoszkiewicz,
  Okada, Jones, Li, King, Marder, and Riedo}}]{szoszkiewicz2007high}
\bibinfo{author}{\bibfnamefont{R.}~\bibnamefont{Szoszkiewicz}},
  \bibinfo{author}{\bibfnamefont{T.}~\bibnamefont{Okada}},
  \bibinfo{author}{\bibfnamefont{S.~C.} \bibnamefont{Jones}},
  \bibinfo{author}{\bibfnamefont{T.-D.} \bibnamefont{Li}},
  \bibinfo{author}{\bibfnamefont{W.~P.} \bibnamefont{King}},
  \bibinfo{author}{\bibfnamefont{S.~R.} \bibnamefont{Marder}},
  \bibnamefont{and} \bibinfo{author}{\bibfnamefont{E.}~\bibnamefont{Riedo}},
  \bibinfo{journal}{Nano Lett.} \textbf{\bibinfo{volume}{7}},
  \bibinfo{pages}{1064} (\bibinfo{year}{2007}).

\bibitem[{\citenamefont{Carroll et~al.}(2013)\citenamefont{Carroll, Giordano,
  Wang, Kodali, Scrimgeour, King, Marder, Riedo, and
  Curtis}}]{carroll2013fabricating}
\bibinfo{author}{\bibfnamefont{K.~M.} \bibnamefont{Carroll}},
  \bibinfo{author}{\bibfnamefont{A.~J.} \bibnamefont{Giordano}},
  \bibinfo{author}{\bibfnamefont{D.}~\bibnamefont{Wang}},
  \bibinfo{author}{\bibfnamefont{V.~K.} \bibnamefont{Kodali}},
  \bibinfo{author}{\bibfnamefont{J.}~\bibnamefont{Scrimgeour}},
  \bibinfo{author}{\bibfnamefont{W.~P.} \bibnamefont{King}},
  \bibinfo{author}{\bibfnamefont{S.~R.} \bibnamefont{Marder}},
  \bibinfo{author}{\bibfnamefont{E.}~\bibnamefont{Riedo}}, \bibnamefont{and}
  \bibinfo{author}{\bibfnamefont{J.~E.} \bibnamefont{Curtis}},
  \bibinfo{journal}{Langmuir} \textbf{\bibinfo{volume}{29}},
  \bibinfo{pages}{8675} (\bibinfo{year}{2013}).

\bibitem[{\citenamefont{Carroll et~al.}(0)\citenamefont{Carroll, Wolf, Knoll,
  Curtis, Zhang, Marder, Riedo, and Duerig}}]{doi:10.1021/acs.langmuir.6b03471}
\bibinfo{author}{\bibfnamefont{K.~M.} \bibnamefont{Carroll}},
  \bibinfo{author}{\bibfnamefont{H.}~\bibnamefont{Wolf}},
  \bibinfo{author}{\bibfnamefont{A.~W.} \bibnamefont{Knoll}},
  \bibinfo{author}{\bibfnamefont{J.~E.} \bibnamefont{Curtis}},
  \bibinfo{author}{\bibfnamefont{Y.}~\bibnamefont{Zhang}},
  \bibinfo{author}{\bibfnamefont{S.~R.} \bibnamefont{Marder}},
  \bibinfo{author}{\bibfnamefont{E.}~\bibnamefont{Riedo}}, \bibnamefont{and}
  \bibinfo{author}{\bibfnamefont{U.}~\bibnamefont{Duerig}},
  \bibinfo{journal}{Langmuir} \textbf{\bibinfo{volume}{0}},
  \bibinfo{pages}{null} (\bibinfo{year}{0}),
  \eprint{http://dx.doi.org/10.1021/acs.langmuir.6b03471},
  \urlprefix\url{http://dx.doi.org/10.1021/acs.langmuir.6b03471}.

\bibitem[{\citenamefont{Wei et~al.}(2010)\citenamefont{Wei, Wang, Kim, Kim, Hu,
  Yakes, Laracuente, Dai, Marder, Berger et~al.}}]{wei2010nanoscale}
\bibinfo{author}{\bibfnamefont{Z.}~\bibnamefont{Wei}},
  \bibinfo{author}{\bibfnamefont{D.}~\bibnamefont{Wang}},
  \bibinfo{author}{\bibfnamefont{S.}~\bibnamefont{Kim}},
  \bibinfo{author}{\bibfnamefont{S.-Y.} \bibnamefont{Kim}},
  \bibinfo{author}{\bibfnamefont{Y.}~\bibnamefont{Hu}},
  \bibinfo{author}{\bibfnamefont{M.~K.} \bibnamefont{Yakes}},
  \bibinfo{author}{\bibfnamefont{A.~R.} \bibnamefont{Laracuente}},
  \bibinfo{author}{\bibfnamefont{Z.}~\bibnamefont{Dai}},
  \bibinfo{author}{\bibfnamefont{S.~R.} \bibnamefont{Marder}},
  \bibinfo{author}{\bibfnamefont{C.}~\bibnamefont{Berger}},
  \bibnamefont{et~al.}, \bibinfo{journal}{Science}
  \textbf{\bibinfo{volume}{328}}, \bibinfo{pages}{1373} (\bibinfo{year}{2010}).

\bibitem[{\citenamefont{Shaw et~al.}(2013)\citenamefont{Shaw, Stavrinou, and
  Anthopoulos}}]{ADMA:ADMA201202877}
\bibinfo{author}{\bibfnamefont{J.~E.} \bibnamefont{Shaw}},
  \bibinfo{author}{\bibfnamefont{P.~N.} \bibnamefont{Stavrinou}},
  \bibnamefont{and} \bibinfo{author}{\bibfnamefont{T.~D.}
  \bibnamefont{Anthopoulos}}, \bibinfo{journal}{Adv. Mater.}
  \textbf{\bibinfo{volume}{25}}, \bibinfo{pages}{552} (\bibinfo{year}{2013}),
  ISSN \bibinfo{issn}{1521-4095},
  \urlprefix\url{http://dx.doi.org/10.1002/adma.201202877}.

\bibitem[{\citenamefont{Albisetti et~al.}(2016)\citenamefont{Albisetti, Petti,
  Pancaldi, Madami, Tacchi, Curtis, King, Papp, Csaba, Porod
  et~al.}}]{albisetti2016nanopatterning}
\bibinfo{author}{\bibfnamefont{E.}~\bibnamefont{Albisetti}},
  \bibinfo{author}{\bibfnamefont{D.}~\bibnamefont{Petti}},
  \bibinfo{author}{\bibfnamefont{M.}~\bibnamefont{Pancaldi}},
  \bibinfo{author}{\bibfnamefont{M.}~\bibnamefont{Madami}},
  \bibinfo{author}{\bibfnamefont{S.}~\bibnamefont{Tacchi}},
  \bibinfo{author}{\bibfnamefont{J.}~\bibnamefont{Curtis}},
  \bibinfo{author}{\bibfnamefont{W.}~\bibnamefont{King}},
  \bibinfo{author}{\bibfnamefont{A.}~\bibnamefont{Papp}},
  \bibinfo{author}{\bibfnamefont{G.}~\bibnamefont{Csaba}},
  \bibinfo{author}{\bibfnamefont{W.}~\bibnamefont{Porod}},
  \bibnamefont{et~al.}, \bibinfo{journal}{Nat. Nanotechnol.}
  \textbf{\bibinfo{volume}{11}}, \bibinfo{pages}{545} (\bibinfo{year}{2016}).

\bibitem[{\citenamefont{Kim et~al.}(2011)\citenamefont{Kim, Bastani, Lu, King,
  Marder, Sandhage, Gruverman, Riedo, and Bassiri-Gharb}}]{kim2011direct}
\bibinfo{author}{\bibfnamefont{S.}~\bibnamefont{Kim}},
  \bibinfo{author}{\bibfnamefont{Y.}~\bibnamefont{Bastani}},
  \bibinfo{author}{\bibfnamefont{H.}~\bibnamefont{Lu}},
  \bibinfo{author}{\bibfnamefont{W.~P.} \bibnamefont{King}},
  \bibinfo{author}{\bibfnamefont{S.}~\bibnamefont{Marder}},
  \bibinfo{author}{\bibfnamefont{K.~H.} \bibnamefont{Sandhage}},
  \bibinfo{author}{\bibfnamefont{A.}~\bibnamefont{Gruverman}},
  \bibinfo{author}{\bibfnamefont{E.}~\bibnamefont{Riedo}}, \bibnamefont{and}
  \bibinfo{author}{\bibfnamefont{N.}~\bibnamefont{Bassiri-Gharb}},
  \bibinfo{journal}{Adv. Mater.} \textbf{\bibinfo{volume}{23}},
  \bibinfo{pages}{3786} (\bibinfo{year}{2011}).

\bibitem[{\citenamefont{Basu et~al.}(2004)\citenamefont{Basu, McNamara, and
  Gianchandani}}]{basu2004scanning}
\bibinfo{author}{\bibfnamefont{A.~S.} \bibnamefont{Basu}},
  \bibinfo{author}{\bibfnamefont{S.}~\bibnamefont{McNamara}}, \bibnamefont{and}
  \bibinfo{author}{\bibfnamefont{Y.~B.} \bibnamefont{Gianchandani}},
  \bibinfo{journal}{J. Vac. Sci. Technol. B} \textbf{\bibinfo{volume}{22}},
  \bibinfo{pages}{3217} (\bibinfo{year}{2004}).

\bibitem[{\citenamefont{Hua et~al.}(2006)\citenamefont{Hua, Saxena, King, and
  Henderson}}]{hua2006nanolithography}
\bibinfo{author}{\bibfnamefont{Y.}~\bibnamefont{Hua}},
  \bibinfo{author}{\bibfnamefont{S.}~\bibnamefont{Saxena}},
  \bibinfo{author}{\bibfnamefont{W.~P.} \bibnamefont{King}}, \bibnamefont{and}
  \bibinfo{author}{\bibfnamefont{C.~L.} \bibnamefont{Henderson}}, in
  \emph{\bibinfo{booktitle}{SPIE 31st International Symposium on Advanced
  Lithography}} (\bibinfo{organization}{International Society for Optics and
  Photonics}, \bibinfo{year}{2006}), pp. \bibinfo{pages}{61531G--61531G}.

\bibitem[{\citenamefont{Gotsmann et~al.}(2006)\citenamefont{Gotsmann, Duerig,
  Frommer, and Hawker}}]{gotsmann2006exploiting}
\bibinfo{author}{\bibfnamefont{B.}~\bibnamefont{Gotsmann}},
  \bibinfo{author}{\bibfnamefont{U.}~\bibnamefont{Duerig}},
  \bibinfo{author}{\bibfnamefont{J.}~\bibnamefont{Frommer}}, \bibnamefont{and}
  \bibinfo{author}{\bibfnamefont{C.~J.} \bibnamefont{Hawker}},
  \bibinfo{journal}{Adv. Funct. Mater.} \textbf{\bibinfo{volume}{16}},
  \bibinfo{pages}{1499} (\bibinfo{year}{2006}).

\bibitem[{\citenamefont{Pires et~al.}(2010)\citenamefont{Pires, Hedrick,
  De~Silva, Frommer, Gotsmann, Wolf, Despont, Duerig, and
  Knoll}}]{pires2010nanoscale}
\bibinfo{author}{\bibfnamefont{D.}~\bibnamefont{Pires}},
  \bibinfo{author}{\bibfnamefont{J.~L.} \bibnamefont{Hedrick}},
  \bibinfo{author}{\bibfnamefont{A.}~\bibnamefont{De~Silva}},
  \bibinfo{author}{\bibfnamefont{J.}~\bibnamefont{Frommer}},
  \bibinfo{author}{\bibfnamefont{B.}~\bibnamefont{Gotsmann}},
  \bibinfo{author}{\bibfnamefont{H.}~\bibnamefont{Wolf}},
  \bibinfo{author}{\bibfnamefont{M.}~\bibnamefont{Despont}},
  \bibinfo{author}{\bibfnamefont{U.}~\bibnamefont{Duerig}}, \bibnamefont{and}
  \bibinfo{author}{\bibfnamefont{A.~W.} \bibnamefont{Knoll}},
  \bibinfo{journal}{Science} \textbf{\bibinfo{volume}{328}},
  \bibinfo{pages}{732} (\bibinfo{year}{2010}).

\bibitem[{\citenamefont{Knoll et~al.}(2010)\citenamefont{Knoll, Pires,
  Coulembier, Dubois, Hedrick, Frommer, and Duerig}}]{knoll2010probe}
\bibinfo{author}{\bibfnamefont{A.~W.} \bibnamefont{Knoll}},
  \bibinfo{author}{\bibfnamefont{D.}~\bibnamefont{Pires}},
  \bibinfo{author}{\bibfnamefont{O.}~\bibnamefont{Coulembier}},
  \bibinfo{author}{\bibfnamefont{P.}~\bibnamefont{Dubois}},
  \bibinfo{author}{\bibfnamefont{J.~L.} \bibnamefont{Hedrick}},
  \bibinfo{author}{\bibfnamefont{J.}~\bibnamefont{Frommer}}, \bibnamefont{and}
  \bibinfo{author}{\bibfnamefont{U.}~\bibnamefont{Duerig}},
  \bibinfo{journal}{Adv. Mater.} \textbf{\bibinfo{volume}{22}},
  \bibinfo{pages}{3361} (\bibinfo{year}{2010}).

\bibitem[{\citenamefont{Lee et~al.}(2007)\citenamefont{Lee, Wright, Abel,
  Sunden, Marchenkov, Graham, and King}}]{lee2007thermal}
\bibinfo{author}{\bibfnamefont{J.}~\bibnamefont{Lee}},
  \bibinfo{author}{\bibfnamefont{T.~L.} \bibnamefont{Wright}},
  \bibinfo{author}{\bibfnamefont{M.~R.} \bibnamefont{Abel}},
  \bibinfo{author}{\bibfnamefont{E.~O.} \bibnamefont{Sunden}},
  \bibinfo{author}{\bibfnamefont{A.}~\bibnamefont{Marchenkov}},
  \bibinfo{author}{\bibfnamefont{S.}~\bibnamefont{Graham}}, \bibnamefont{and}
  \bibinfo{author}{\bibfnamefont{W.~P.} \bibnamefont{King}},
  \bibinfo{journal}{J. Appl. Phys.} \textbf{\bibinfo{volume}{101}},
  \bibinfo{pages}{014906} (\bibinfo{year}{2007}).

\bibitem[{\citenamefont{Nelson and King}(2007)}]{nelson2007temperature}
\bibinfo{author}{\bibfnamefont{B.~A.} \bibnamefont{Nelson}} \bibnamefont{and}
  \bibinfo{author}{\bibfnamefont{W.}~\bibnamefont{King}},
  \bibinfo{journal}{Sensor. Actuat. A-Phys.} \textbf{\bibinfo{volume}{140}},
  \bibinfo{pages}{51} (\bibinfo{year}{2007}).

\bibitem[{\citenamefont{Despont et~al.}(2000)\citenamefont{Despont, Brugger,
  Drechsler, D{\"u}rig, H{\"a}berle, Lutwyche, Rothuizen, Stutz, Widmer, Binnig
  et~al.}}]{despont2000vlsi}
\bibinfo{author}{\bibfnamefont{M.}~\bibnamefont{Despont}},
  \bibinfo{author}{\bibfnamefont{J.}~\bibnamefont{Brugger}},
  \bibinfo{author}{\bibfnamefont{U.}~\bibnamefont{Drechsler}},
  \bibinfo{author}{\bibfnamefont{U.}~\bibnamefont{D{\"u}rig}},
  \bibinfo{author}{\bibfnamefont{W.}~\bibnamefont{H{\"a}berle}},
  \bibinfo{author}{\bibfnamefont{M.}~\bibnamefont{Lutwyche}},
  \bibinfo{author}{\bibfnamefont{H.}~\bibnamefont{Rothuizen}},
  \bibinfo{author}{\bibfnamefont{R.}~\bibnamefont{Stutz}},
  \bibinfo{author}{\bibfnamefont{R.}~\bibnamefont{Widmer}},
  \bibinfo{author}{\bibfnamefont{G.}~\bibnamefont{Binnig}},
  \bibnamefont{et~al.}, \bibinfo{journal}{Sensor. Actuat. A-Phys.}
  \textbf{\bibinfo{volume}{80}}, \bibinfo{pages}{100} (\bibinfo{year}{2000}).

\bibitem[{\citenamefont{Drechsler et~al.}(2003)\citenamefont{Drechsler,
  B{\"u}rer, Despont, D{\"u}rig, Gotsmann, Robin, and
  Vettiger}}]{drechsler2003cantilevers}
\bibinfo{author}{\bibfnamefont{U.}~\bibnamefont{Drechsler}},
  \bibinfo{author}{\bibfnamefont{N.}~\bibnamefont{B{\"u}rer}},
  \bibinfo{author}{\bibfnamefont{M.}~\bibnamefont{Despont}},
  \bibinfo{author}{\bibfnamefont{U.}~\bibnamefont{D{\"u}rig}},
  \bibinfo{author}{\bibfnamefont{B.}~\bibnamefont{Gotsmann}},
  \bibinfo{author}{\bibfnamefont{F.}~\bibnamefont{Robin}}, \bibnamefont{and}
  \bibinfo{author}{\bibfnamefont{P.}~\bibnamefont{Vettiger}},
  \bibinfo{journal}{Microelectron. Eng.} \textbf{\bibinfo{volume}{67}},
  \bibinfo{pages}{397} (\bibinfo{year}{2003}).

\bibitem[{\citenamefont{D{\"u}rig}(2005)}]{durig2005fundamentals}
\bibinfo{author}{\bibfnamefont{U.}~\bibnamefont{D{\"u}rig}},
  \bibinfo{journal}{J. Appl. Phys.} \textbf{\bibinfo{volume}{98}},
  \bibinfo{pages}{044906} (\bibinfo{year}{2005}).

\bibitem[{\citenamefont{Asheghi et~al.}(2002)\citenamefont{Asheghi,
  Kurabayashi, Kasnavi, and Goodson}}]{asheghi2002thermal}
\bibinfo{author}{\bibfnamefont{M.}~\bibnamefont{Asheghi}},
  \bibinfo{author}{\bibfnamefont{K.}~\bibnamefont{Kurabayashi}},
  \bibinfo{author}{\bibfnamefont{R.}~\bibnamefont{Kasnavi}}, \bibnamefont{and}
  \bibinfo{author}{\bibfnamefont{K.}~\bibnamefont{Goodson}},
  \bibinfo{journal}{J. Appl. Phys.} \textbf{\bibinfo{volume}{91}},
  \bibinfo{pages}{5079} (\bibinfo{year}{2002}).

\bibitem[{\citenamefont{Fulkerson et~al.}(1968)\citenamefont{Fulkerson, Moore,
  Williams, Graves, and McElroy}}]{fulkerson1968thermal}
\bibinfo{author}{\bibfnamefont{W.}~\bibnamefont{Fulkerson}},
  \bibinfo{author}{\bibfnamefont{J.}~\bibnamefont{Moore}},
  \bibinfo{author}{\bibfnamefont{R.}~\bibnamefont{Williams}},
  \bibinfo{author}{\bibfnamefont{R.}~\bibnamefont{Graves}}, \bibnamefont{and}
  \bibinfo{author}{\bibfnamefont{D.}~\bibnamefont{McElroy}},
  \bibinfo{journal}{Phys. Rev.} \textbf{\bibinfo{volume}{167}},
  \bibinfo{pages}{765} (\bibinfo{year}{1968}).

\bibitem[{\citenamefont{Stranz et~al.}(2013)\citenamefont{Stranz, K{\"a}hler,
  Waag, and Peiner}}]{stranz2013thermoelectric}
\bibinfo{author}{\bibfnamefont{A.}~\bibnamefont{Stranz}},
  \bibinfo{author}{\bibfnamefont{J.}~\bibnamefont{K{\"a}hler}},
  \bibinfo{author}{\bibfnamefont{A.}~\bibnamefont{Waag}}, \bibnamefont{and}
  \bibinfo{author}{\bibfnamefont{E.}~\bibnamefont{Peiner}},
  \bibinfo{journal}{J. Electron. Mater.} \textbf{\bibinfo{volume}{42}},
  \bibinfo{pages}{2381} (\bibinfo{year}{2013}).

\bibitem[{\citenamefont{Chang}(1963)}]{chang1963concentration}
\bibinfo{author}{\bibfnamefont{J.}~\bibnamefont{Chang}}, \bibinfo{journal}{IEEE
  T. Electron. Dev.} \textbf{\bibinfo{volume}{10}}, \bibinfo{pages}{357}
  (\bibinfo{year}{1963}).

\bibitem[{\citenamefont{Thai}(1970)}]{thai1970concentration}
\bibinfo{author}{\bibfnamefont{N.}~\bibnamefont{Thai}}, \bibinfo{journal}{J.
  Appl. Phys.} \textbf{\bibinfo{volume}{41}}, \bibinfo{pages}{2859}
  (\bibinfo{year}{1970}).

\bibitem[{\citenamefont{Altermatt et~al.}(1999)\citenamefont{Altermatt, Schenk,
  Heiser, and Green}}]{altermatt1999influence}
\bibinfo{author}{\bibfnamefont{P.}~\bibnamefont{Altermatt}},
  \bibinfo{author}{\bibfnamefont{A.}~\bibnamefont{Schenk}},
  \bibinfo{author}{\bibfnamefont{G.}~\bibnamefont{Heiser}}, \bibnamefont{and}
  \bibinfo{author}{\bibfnamefont{M.}~\bibnamefont{Green}}, in
  \emph{\bibinfo{booktitle}{Int. Photovoltaic Sci. Eng. Conf., Sapporo}}
  (\bibinfo{year}{1999}), vol.~\bibinfo{volume}{11}, p. \bibinfo{pages}{719}.

\bibitem[{\citenamefont{Berggren and Sernelius}(1981)}]{berggren1981band}
\bibinfo{author}{\bibfnamefont{K.-F.} \bibnamefont{Berggren}} \bibnamefont{and}
  \bibinfo{author}{\bibfnamefont{B.~E.} \bibnamefont{Sernelius}},
  \bibinfo{journal}{Phys. Rev. B} \textbf{\bibinfo{volume}{24}},
  \bibinfo{pages}{1971} (\bibinfo{year}{1981}).

\bibitem[{\citenamefont{Chapman et~al.}(1963)\citenamefont{Chapman, Tufte,
  Zook, and Long}}]{chapman1963electrical}
\bibinfo{author}{\bibfnamefont{P.}~\bibnamefont{Chapman}},
  \bibinfo{author}{\bibfnamefont{O.}~\bibnamefont{Tufte}},
  \bibinfo{author}{\bibfnamefont{J.~D.} \bibnamefont{Zook}}, \bibnamefont{and}
  \bibinfo{author}{\bibfnamefont{D.}~\bibnamefont{Long}}, \bibinfo{journal}{J.
  Appl. Phys.} \textbf{\bibinfo{volume}{34}}, \bibinfo{pages}{3291}
  (\bibinfo{year}{1963}).

\bibitem[{\citenamefont{Abel et~al.}(2007)\citenamefont{Abel, Wright, King, and
  Graham}}]{abel2007thermal}
\bibinfo{author}{\bibfnamefont{M.~R.} \bibnamefont{Abel}},
  \bibinfo{author}{\bibfnamefont{T.~L.} \bibnamefont{Wright}},
  \bibinfo{author}{\bibfnamefont{W.~P.} \bibnamefont{King}}, \bibnamefont{and}
  \bibinfo{author}{\bibfnamefont{S.}~\bibnamefont{Graham}},
  \bibinfo{journal}{IEEE T. Compon. Pack. T.} \textbf{\bibinfo{volume}{30}},
  \bibinfo{pages}{200} (\bibinfo{year}{2007}).

\bibitem[{\citenamefont{L{\"o}rtscher et~al.}(2013)\citenamefont{L{\"o}rtscher,
  Widmer, and Gotsmann}}]{lortscher2013next}
\bibinfo{author}{\bibfnamefont{E.}~\bibnamefont{L{\"o}rtscher}},
  \bibinfo{author}{\bibfnamefont{D.}~\bibnamefont{Widmer}}, \bibnamefont{and}
  \bibinfo{author}{\bibfnamefont{B.}~\bibnamefont{Gotsmann}},
  \bibinfo{journal}{Nanoscale} \textbf{\bibinfo{volume}{5}},
  \bibinfo{pages}{10542} (\bibinfo{year}{2013}).

\bibitem[{\citenamefont{Hart et~al.}(1970)\citenamefont{Hart, Aggarwal, and
  Lax}}]{PhysRevB.1.638}
\bibinfo{author}{\bibfnamefont{T.~R.} \bibnamefont{Hart}},
  \bibinfo{author}{\bibfnamefont{R.~L.} \bibnamefont{Aggarwal}},
  \bibnamefont{and} \bibinfo{author}{\bibfnamefont{B.}~\bibnamefont{Lax}},
  \bibinfo{journal}{Phys. Rev. B} \textbf{\bibinfo{volume}{1}},
  \bibinfo{pages}{638} (\bibinfo{year}{1970}),
  \urlprefix\url{http://link.aps.org/doi/10.1103/PhysRevB.1.638}.

\bibitem[{\citenamefont{Ganesan et~al.}(1970)\citenamefont{Ganesan, Maradudin,
  and Oitmaa}}]{ganesan1970lattice}
\bibinfo{author}{\bibfnamefont{S.}~\bibnamefont{Ganesan}},
  \bibinfo{author}{\bibfnamefont{A.}~\bibnamefont{Maradudin}},
  \bibnamefont{and} \bibinfo{author}{\bibfnamefont{J.}~\bibnamefont{Oitmaa}},
  \bibinfo{journal}{Ann. Phys.} \textbf{\bibinfo{volume}{56}},
  \bibinfo{pages}{556} (\bibinfo{year}{1970}).

\bibitem[{\citenamefont{Chui et~al.}(1999)\citenamefont{Chui, Asheghi, Ju,
  Goodson, Kenny, and Mamin}}]{chui1999intrinsic}
\bibinfo{author}{\bibfnamefont{B.}~\bibnamefont{Chui}},
  \bibinfo{author}{\bibfnamefont{M.}~\bibnamefont{Asheghi}},
  \bibinfo{author}{\bibfnamefont{Y.}~\bibnamefont{Ju}},
  \bibinfo{author}{\bibfnamefont{K.}~\bibnamefont{Goodson}},
  \bibinfo{author}{\bibfnamefont{T.}~\bibnamefont{Kenny}}, \bibnamefont{and}
  \bibinfo{author}{\bibfnamefont{H.}~\bibnamefont{Mamin}},
  \bibinfo{journal}{Microscale Therm. Eng.} \textbf{\bibinfo{volume}{3}},
  \bibinfo{pages}{217} (\bibinfo{year}{1999}).

\bibitem[{\citenamefont{Logg et~al.}(2012)\citenamefont{Logg, Mardal, and
  Wells}}]{logg2012automated}
\bibinfo{author}{\bibfnamefont{A.}~\bibnamefont{Logg}},
  \bibinfo{author}{\bibfnamefont{K.-A.} \bibnamefont{Mardal}},
  \bibnamefont{and} \bibinfo{author}{\bibfnamefont{G.}~\bibnamefont{Wells}},
  \emph{\bibinfo{title}{Automated solution of differential equations by the
  finite element method: The FEniCS book}}, vol.~\bibinfo{volume}{84}
  (\bibinfo{publisher}{Springer Science \& Business Media},
  \bibinfo{year}{2012}).

\bibitem[{\citenamefont{Aln{\ae}s et~al.}(2015)\citenamefont{Aln{\ae}s,
  Blechta, Hake, Johansson, Kehlet, Logg, Richardson, Ring, Rognes, and
  Wells}}]{alnaes2015fenics}
\bibinfo{author}{\bibfnamefont{M.}~\bibnamefont{Aln{\ae}s}},
  \bibinfo{author}{\bibfnamefont{J.}~\bibnamefont{Blechta}},
  \bibinfo{author}{\bibfnamefont{J.}~\bibnamefont{Hake}},
  \bibinfo{author}{\bibfnamefont{A.}~\bibnamefont{Johansson}},
  \bibinfo{author}{\bibfnamefont{B.}~\bibnamefont{Kehlet}},
  \bibinfo{author}{\bibfnamefont{A.}~\bibnamefont{Logg}},
  \bibinfo{author}{\bibfnamefont{C.}~\bibnamefont{Richardson}},
  \bibinfo{author}{\bibfnamefont{J.}~\bibnamefont{Ring}},
  \bibinfo{author}{\bibfnamefont{M.~E.} \bibnamefont{Rognes}},
  \bibnamefont{and} \bibinfo{author}{\bibfnamefont{G.~N.} \bibnamefont{Wells}},
  \bibinfo{journal}{Archive of Numerical Software}
  \textbf{\bibinfo{volume}{3}}, \bibinfo{pages}{9} (\bibinfo{year}{2015}).

\bibitem[{lmf(2014)}]{lmfit}
\emph{\bibinfo{title}{LMFIT: Non-Linear Least-Square Minimization and
  Curve-Fitting for Python}} (\bibinfo{year}{2014}),
  \urlprefix\url{http://dx.doi.org/10.5281/zenodo.11813}.

\bibitem[{\citenamefont{Yovanovich}(1970)}]{yovanovich1970temperature}
\bibinfo{author}{\bibfnamefont{M.~M.} \bibnamefont{Yovanovich}},
  \bibinfo{journal}{J. Compos. Mater.} \textbf{\bibinfo{volume}{4}},
  \bibinfo{pages}{567} (\bibinfo{year}{1970}).

\bibitem[{\citenamefont{Freeman and Lowther}(1989)}]{freeman1989open}
\bibinfo{author}{\bibfnamefont{E.}~\bibnamefont{Freeman}} \bibnamefont{and}
  \bibinfo{author}{\bibfnamefont{D.}~\bibnamefont{Lowther}},
  \bibinfo{journal}{IEEE T. Magn.} \textbf{\bibinfo{volume}{25}},
  \bibinfo{pages}{4135} (\bibinfo{year}{1989}).

\bibitem[{\citenamefont{Kadoya et~al.}(1985)\citenamefont{Kadoya, Matsunaga,
  and Nagashima}}]{kadoya1985viscosity}
\bibinfo{author}{\bibfnamefont{K.}~\bibnamefont{Kadoya}},
  \bibinfo{author}{\bibfnamefont{N.}~\bibnamefont{Matsunaga}},
  \bibnamefont{and}
  \bibinfo{author}{\bibfnamefont{A.}~\bibnamefont{Nagashima}},
  \bibinfo{journal}{J. Phys. Chem. Ref. Data} \textbf{\bibinfo{volume}{14}},
  \bibinfo{pages}{947} (\bibinfo{year}{1985}).

\end{thebibliography}

\end{document}